\documentclass[procedia]{easychair}
\usepackage{doc}
\usepackage{makeidx}

\usepackage{amssymb}
\usepackage{amsthm}
\usepackage{amsmath}
\usepackage{bm}% bold math
\usepackage{amssymb, amsmath}

\def\procediaConference{International Conference on Computational Science (ICCS 2017)}

\makeindex

\title{Brownian dynamics simulations to explore experimental microsphere diffusion
with optical tweezers.}

\titlerunning{Brownian dynamics simulations to explore microsphere diffusion\ldots}

\author{
    Manuel Pancorbo\inst{1}%\thanks{Designed and implemented the class style}
\and
    Miguel A. Rubio\inst{2}%\thanks{Did numerous tests and provided a lot of suggestions}
\and
    P. Dom\'{i}nguez-Garc\'{i}a \inst{1}\\
}

\institute{
  Dep. F\'isica Interdisciplinar, Universidad Nacional de Educaci\'on a Distancia (UNED), Senda del Rey 9, Madrid 28040, Spain.
\and
  Dep. F\'isica Fundamental, Universidad Nacional de Educaci\'on a Distancia (UNED), Senda del Rey 9, Madrid 28040, Spain.
 }
\authorrunning{Pancorbo, Rubio and Dom\'{i}nguez-Garc\'{i}a}

\begin{document}

\maketitle


\begin{abstract}

We develop two-dimensional Brownian dynamics simulations to examine 
the motion of disks under thermal fluctuations and Hookean forces. 
Our simulations are designed to be 
experimental-like, since the experimental conditions define the available 
time-scales which characterize the solution of Langevin equations. 
To define the fluid model and methodology, 
we explain the basics of the theory of Brownian motion applicable to
quasi-twodimensional diffusion of optically-trapped microspheres. 
Using the data produced by the simulations, 
we propose an alternative methodology to calculate diffusion coefficients.
We obtain that, using typical input parameters in video-microscopy experiments, 
the averaged values of the diffusion coefficient differ from the theoretical one
less than a 1\%.

\end{abstract}

%\setcounter{tocdepth}{2}
%{\small
%\tableofcontents}

%\section{To mention}
%
%Processing in EasyChair - number of pages.
%
%Examples of how EasyChair processes papers. Caveats (replacement of EC
%class, errors).

%------------------------------------------------------------------------------
\section{Introduction}
%\label{sect:introduction}
Brownian motion, i.e., the random movement of objects immersed in a fluid, 
was theoretically described by Einstein more than a century ago \cite{Xin2016}
from a microscopic perspective, 
demonstrating the molecular structure of the fluid \cite{einstein_motion_1905}. 
The Einstein's classical approach neglected hydrodynamics memory and inertia effects, 
since they appear at very short-time scales, 
something experimentally available only very recently \cite{li_t_brownian_2013}.
This assumption theoretically implies that the particle velocity cannot be
defined and the trajectories of Brownian particles are fractal \cite{Pusey2011}.
Therefore, the study of Brownian motion is determined by the available experimental set-up,
which defines the detected time-resolution of the stochastic jumps.

The standard experimental methodology to study Brownian motion is to mix
a small concentration of micro-nanospheres with a certain fluid.
The suspension sample is deposited into a glass cell which is inserted 
in an optical instrument, such as a video-microscopy or an interferometry set-up, 
where the trajectories of the beads can be recorded for ulterior analysis. 
A common practice to facilitate the study of particles' motion is using optical tweezers \cite{ashkin_applications_1980}. 
This technique exerts a restoring force under the object, 
allowing the experimentalist to move the particle inside the fluid in a quasi-twodimensional plane. 
Many optical tweezers set-ups allow to trap several objects, 
but single-particle tracking is usually employed to improve spatial and temporal resolution.

During the last decades, optical traps have permitted to develop 
a wide variety of experiments in colloidal motion.
To cite some examples: about the effects 
caused by confinement \cite{jeney_anisotropic_2008}, 
the hydrodynamic interaction between particles \cite{dufresne_hydrodynamic_2000}, 
discovering resonances from hydrodynamic memory at short-time scales \cite{franosch_resonances_2011}, 
regarding micro-rheology \cite{tassieri_microrheology_2012},
or even to produce Brownian Carnot engines \cite{MartinezCarnot2016}, 
along with many other applications \cite{GrierOptrev2003}.

In spite of the evident benefits of optical tweezers in the research of colloidal physics, 
this experimental methodology generates an external force under the bead which can modify the dynamics of the particle \cite{lukic_motion_2007}.
This force can be also itself modified by the thermo-physical properties of the surrounding complex fluid \cite{Dominguez2016}. 
Therefore, a good dynamical characterization of the external harmonic potential 
detected by stochastic particle motion is needed to correctly measure the values of the diffusion coefficient.

Here, we study by experimental-like computer simulations the diffusion of a single sphere 
observed in a two-dimensional plane under optical tweezers, \textit{i.e.}, 
we investigate the Brownian motion of a disk under harmonic potentials.  
In this work, we show how the solution of the Fokker-Planck equation \cite{Risken1984}
allows us to propose an iterative approach as an alternative methodology 
to calculate the diffusion coefficient of the disk. 
%Version1.1
Our objective in this work is to emulate the dynamics of a trapped single-bead in a 
video-microscopy experiment by means of Brownian dynamics simulations. 
These simulations are designed to be experimental-like using typical input parameters 
but without the limitations which appear in an experimental 
set-up, like image analysis miscalculations or confinement effects in the bead's diffusion.
%End Version1.1

\section{Brownian dynamics simulations}

We develop Brownian dynamics simulations, 
which are a simplification of Stokestian dynamics, 
but neglecting hydrodynamic interactions (HI) between particles \cite{hess_generalizated_1983}.
Our model of colloidal fluid is designed to be compared with video-microscopy (VM) experiments, 
where we can observe real-time motion of the colloids in two dimensions 
and where we can storage the particles' position for defined temporal steps, according to 
the frame-rate of the camera. 
The software simulates a suspension of micro-spheres in a Newtonian fluid
in a 2D or pseudo-2D configuration of sedimented micro-particles \cite{dominguez-garcia_p_single_2013}.
In analogy with a image-based VM lab, 
we are able to change external parameters, expressed in physical units, 
such as the concentration of particles in the suspension, 
the viscosity of the fluid, the focal distance, the size of the spheres or the temperature of the bath.

%Version 1.1
Our simulation model has been implemented using an open-source
Java-based software named ``Easy Java/JavaScript Simulations'' (EjsS) \cite{esquembre_easy_2004}, 
allowing to create visual simulations of physical systems based on ordinary differential equations. 
The equations described in this section have been resolved numerically by EjsS 
using a Euler-Richardson algorithm ---alternative 
algorithms are available but they provided the same results. 
This simulation methodology has been successfully 
developed and tested in more complex systems 
composed of many Brownian particles under different internal and external forces \cite{dominguez-garcia_microrheological_2012, dominguez-garcia_p_single_2013}.
%end Version 1.1

%model
Theoretically, the movement of particles under thermal fluctuations is studied by means of the Langevin equation \cite{Langevin1908}, 
which is the Newton's second law equation including a stochastic force $\textbf{F}_{\text{B}}$:
\begin{equation}
m \dot{\textbf{v}} = \textbf{F}_{\text{H}}+\textbf{F}_{\text{B}}+\textbf{F}_{\text{E}}+\textbf{F}_{\text{D}} \label{langevin}
\end{equation}
for a sphere of radius $a$ and mass $m$ immersed in a medium of viscosity $\eta$ and density $\rho$.
In eq. (\ref{langevin}), $\textbf{v}$ is the velocity vector, $\dot{\textbf{v}} \equiv d\textbf{v}/dt$, 
$\textbf{F}_{\text{H}}$ are the hydrodynamic forces, $\textbf{F}_{\text{E}}$ are external 
forces over the particles in the fluid,
and $\textbf{F}_{\text{D}}$ are hard-disk forces that avoid the particles from overlapping. 

We neglect full hydrodynamic interactions, $\textbf{\.{v}} \sim 0$, and, therefore, 
the hydrodynamic contribution is reduced to $\textbf{F}_{\text{H}} = -\gamma\, \textbf{v}$, \textit{i.e.},
the Stokes drag of an isolated particle, where $\gamma = 6\pi\eta a$ is the friction coefficient.
%Version 1.1
This inertial term can be neglected by evaluating the Reynolds number, Re, which compares inertial and viscous forces. For a bead of radius $a$ immersed in a fluid of viscosity $\eta$
and density $\rho$, we have $\text{Re}= \rho v a / \eta$. 
For micro-particles in water-like fluids, Re is low enough, allowing us to exclude hydrodynamic interactions
when the experimental time-scale is not very low ---in the order of the microsecond.\cite{huang_r_direct_2011} 
%End Version 1.1
When using HI, the lubrication forces prevent the particles to overlap. 
If not, hard-disk forces have to be included.
In this work, we use single-particle configuration, and the hard-disk forces are not necessary.

The Brownian or stochastic force is characterized by $\langle \textbf{F}_{\text{B}} \rangle =0$ and by
$\langle \textbf{F}_{\text{B}}(0)\,\textbf{F}_{\text{B}}(t) \rangle = 2 k_B T \gamma \,\bm\delta(t)$
where $k_B$ is the Boltzmann's constant and $\bm\delta(t)$ is the unit tensor \cite{larson_structure_1999}. 
To implement this force, we use a random vector $\mathbf{n}$ with values in the interval $[-1,1]$ generated
by the Box-Muller transformation \cite{Box1959}. Then, we have:
\begin{equation}
 \textbf{F}_{\text{B}}= \sqrt{2d k_B T \gamma / dt} \,\mathbf{n} \label{ruido2}
\end{equation}
where here $d$ is the dimension, and $dt$ will be the time step in the simulations. 

%Harmonic force
Regarding the external forces, 
we use a restoring Hooke-like one: $\textbf{F}_{\text{E}} = -\kappa \textbf{r}$,
where $\kappa$ is the trap stiffness and 
assuming that the trapped disk is centered in its initial position.
We do not include any difference between $x$ and $y$ coordinates, 
and we can define the harmonic potential in one-dimension $x$ as 
$U_\kappa = (1/2)\,\kappa x^ 2$.
This potential is the harmonic approximation 
for a trapping potential generated by the focused laser beam of the optical tweezers, 
which is valid for the central region of the potential well \cite{Svoboda2002, Richardson08}.

Under this conditions, we have the following overdamped Langevin equation for the trapped disk:
\begin{equation}
 \dot{\textbf{r}} = \gamma^{-1}\,\left(-\kappa_r\,\textbf{r}+\textbf{F}_{\text{B}}\right) \label{langevin2}
\end{equation}
where the stochastic term is given by eq. (\ref{ruido2}).

Finally, we make the Stokes-Einstein relation for self-diffusion coefficient explicit \cite{einstein_motion_1906}:
\begin{equation}
D = k_B T \gamma^{-1}\label{stokes-einstein}
\end{equation}
This diffusion coefficient, $D$, is the quantity we want to obtain by analyzing the data of 
the simulations. 
The most simple version of micro-rheology consists in estimating the value of the fluid's viscosity, $\eta$, 
by calculating $D$ from the analysis of particle motion in the fluid.

%easyjava

\section{Harmonic potentials in stochastic motion}

The solution of the complete set of Langevin equations for a spherical particle in harmonic potentials with no-slip
boundaries in a Newtonian fluid and with hydrodynamic effects \cite{clercx_h.j.h._brownian_1992} 
only depends on several timescales: $\tau_f \equiv \rho_f a^2/\eta$, $\tau^{*}_p \equiv m^{*}/\gamma$ and $\tau_\kappa \equiv \gamma/\kappa $.
where $\rho_f$ is the density of the fluid, $m_p$ and the mass of the particle and 
$m^{*} \equiv m_p+m_f/2$ is a modified mass influenced by hydrodynamics, 
where $m_f$ is the mass of the displaced fluid. 
The two first time scales are related to fluid vortex propagation and inertial time-scales, 
whereas $\tau_\kappa$ measures the ratio between the Stokes friction coefficient and the optical trap constant, $\kappa$.

%msd
The most applied statistic magnitude in the analysis of Brownian motion is the mean-square displacement (MSD), 
defined by:
\begin{equation}
\textrm{MSD}(t) \equiv \langle\Delta \textbf{r}^2(t)\rangle \equiv \langle (\textbf{r}(t)-\textbf{r}(0))^2\rangle \label{msd}
\end{equation}
The MSD for a micro-sized particle immersed in a Newtonian fluid, 
in time scales lower than $\tau_\kappa$, behaves as $\textrm{MSD}(t) = 2dD t$, where $d$ is the dimension on the MSD.
This expression defines the diffusive behavior and allows to calculate the diffusion coefficient $D$.
At even lower timescales, the ballistic regime is predominant initiating a temporal power-law behavior, 
where $\textrm{MSD}(t) \sim t^{2}$. 
At higher times, when $t\sim \tau_\kappa$, the optical trap is dominant, 
and the MSD shows a plateau equal to $\textrm{MSD}(t > \tau_\kappa)=2k_B T/\kappa$. 
The one-dimensional MSD of a harmonically trapped bead in a Newtonian fluid, without inertia effects, is \cite{doi_theory_1986}:
\begin{equation}
\langle \Delta x^2(t) \rangle  = \frac{2 k_B T}{\kappa}\,\left(1-e^{-\kappa t/\gamma}\right) \label{Chandrasekhar}
\end{equation}
A standard method to obtain the self-diffusion coefficient is to fit eq. (\ref{Chandrasekhar}) to the MSD data calculated from the disk's positions.
 
%gaussian propagator
An important statistical quantity is the probability distribution of the particles' jumps, $\Delta \textbf{r}$, 
also called Gaussian propagator or van Hove autocorrelation function \cite{Hofling2013}:
\begin{equation}
\rho_D (\Delta \textbf{r}, \tau)= \frac{1}{\left(4 \pi D \tau \right)^{d/2}}\,\exp\left(-\frac{\Delta \textbf{r}^2}{4D\tau}\right)\label{propagador}
\end{equation}
for a fixed and constant time-lapse $\tau$ between particles' jumps in $d$ dimensions.
If we assume there is no difference in the diffusion process by using $\tau$ (lapse time) or $t$ (absolute time), 
the moments of the distribution can be obtained from the propagator by 
$
\delta \textbf{r}^n(t) \equiv \left< |\Delta \textbf{r}(t)|^n\right> = \int |\Delta \textbf{r}|^n\,\rho_D(\Delta \textbf{r},t)\,d^d (\Delta \textbf{r}).
$
And then, the MSD is calculated using $n=2$, obtaining $ \delta \textbf{r}^2(t) \equiv \text{MSD}(t) = 2d D t$.

\begin{figure}[tb]
	\begin{centering}
	\includegraphics[width=0.5\textwidth]{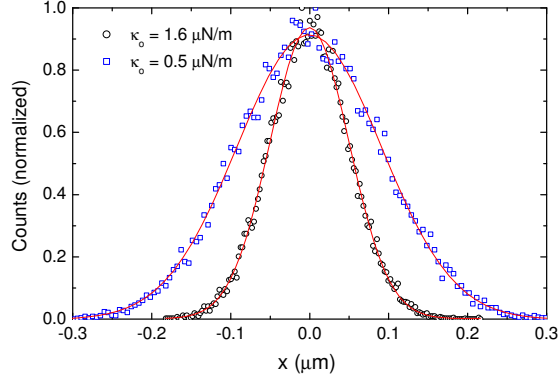}
	\caption{Example of probability density distribution in the $x$ coordinate from the motion of a disk under harmonic potentials in Brownian dynamics simulations
	(diameter $d= 1.9$ $\mu$m, $T=295.5$ K, time-lapse $\tau=2.5$ ms, see Results section). 
	 We show the distribution for two input optical stiffnesses: $\kappa_o = 0.5$ $\mu$N/m (blue points) and $1.6$ $\mu$N/m (red points).
	 The red lines are non-linear fits to Gaussian functions. The value of $\kappa$ can be recovered through the variance of the distributions.}
	\label{fig:rho-k}
	\end{centering}
\end{figure}

%Boltzmann
The diffusion coefficient can be extracted from the one-dimensional variance of the Gaussian distribution (\ref{propagador}) since $\sigma_x^ 2 =  2D\tau$.
A similar approach can be applied to the trap stiffness through Boltzmann statistics \cite{Florin1998} on the particles' positions.
The average motion of a trapped particle can be described by means of the probability density distribution $\rho(x,t)$ which obeys the Fokker-Planck equation \cite{Risken1984,Harada1996}.
This can be written \cite{Davis2007} in a simple one-dimensional form as 
$d\rho(x)/dx = (k_B T)^{-1}\,F(x)\rho(x)$. 
If we use an optical trap modeled as a restoring force with spring constant $\kappa$,
the solution is a Gaussian function on the coordinate $x$: 
\begin{equation}
\rho_\kappa (x) = \left(\frac{\kappa}{2\pi k_B T}\right)^{1/2} \exp\left(-\frac{1}{2} \frac{\kappa x^2}{k_B T}\right)  \label{Boltzmann}
\end{equation}
The variance of the distribution allows to obtain the stiffness of the trap by $\sigma_\kappa^ 2 = k_B T/\kappa$. An example
of these one-dimensional distributions can be seen in Fig. \ref{fig:rho-k}.

%fokker-plank
In a more general case of stochastic motion confined by harmonic potentials, 
the solution of the Fokker-Plank equation leads to a more complicated distribution \cite{Risken1984}. 
In that situation, the probability of transition of a trapped colloidal particle from 
the position $x_0$ to $x$ in lapse-time $\tau$ is:
\begin{equation}
  \rho_{\text{FP}}(x_0, x, \tau) =  \frac{1}{\sqrt{2\pi \alpha(\tau)} }
    \exp{\left[ - \frac{\left(x - x_0 e^{-\lambda \tau}\right)^2}{2 \alpha(\tau)} \right] }
    \label{eq:main}
\end{equation}
where:
\begin{subequations}
 \begin{align}
\lambda &\equiv \kappa / \gamma \label{lambda}\\
\alpha(\tau) &\equiv \frac{k_B T}{\kappa}\left(1 - e^{-2\lambda \tau}\right) \label{alpha}
\end{align}
\end{subequations}

Note that $\lambda$ can be written in terms of the diffusion coefficient as $\lambda = \kappa D/k_B T$
by using eq. (\ref{stokes-einstein}) and (\ref{lambda}).
For short values of the time-step, $\tau \ll 1/\lambda$, 
this distribution is identical to the standard Gaussian propagator, (\ref{propagador}).
On the other hand, for long values of the temporal step, it approximates to distribution of particle positions, (\ref{Boltzmann}).    
For intermediate times, the distribution depends of a memory factor which appears in the initial position for every jump of the particle. 
To obtain that memory factor, $e^{-\lambda \tau}$, first we need to know the diffusion coefficient, $D$, 
which is the quantity we want to estimate from the disk's movement.
It is important to observe that, when $\tau \ll 1/\lambda$, we obtain $\alpha(t) = 2Dt$.
Eq. (\ref{lambda}) only differs from the standard MSD, eq. (\ref{Chandrasekhar}), by a factor 2 in the spring constant ($\kappa \rightarrow \kappa/2$).
This will allow us to define a \textit{relaxed} MSD through the temporal jumps with memory effect, 
which we will identify with eq. (\ref{lambda}).

\section{Results}

We use the model previously explained, where the disk's 
positions are obtained following the overdamped Langevin equation, eq. (\ref{langevin2}),
to analyze the diffusion of a trapped simulated disk, always under standard experimental conditions,
when we the increase the stiffness of applied the restoring force.
The input data are those of a typical experiment using commercial optical tweezers: 
trapped particle of diameter $d= 1.9$ $\mu$m, at temperature $T=295.5$ K, 
time-lapse $\tau=2.5$ ms (400 images per second in the video-microscopy set-up), 
during a total time of $50$ s. 
The internal time step in the simulations, $dt$, is fixed to $dt=10^{-4}$ s. 
The typical stiffnesses of the traps are $\kappa \sim 0.02-2$ $\mu$N/m.
The surrounding fluid is water ($\eta=0.95$ mPa.s) and the bead 
is located far enough from the influence of nearby walls \cite{brenner_slow_1961}. 
Using these experimental input values,
we are in the intermediate situation of the Fokker-Planck solution described in the former section.
Indeed, $1/\lambda \sim 9-840$ ms is obtained in the interval of typical $\kappa$, 
not far from the input time-step, $\tau=2.5$ ms.
Here lies the importance of using experimental-like simulations, 
since Brownian motion theoretical explanations depend on the time scale of observation.
Under this conditions, the theoretical diffusion coefficient is $D = 0.239$ $\mu$m$^2$/s according to eq. (\ref{stokes-einstein}).

\begin{figure}[tb]
	\begin{centering}
	\includegraphics[width=0.5\textwidth]{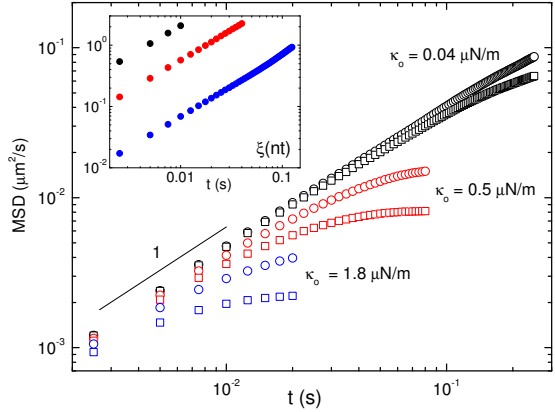}
	\caption{Mean-squared displacements (MSDs) from two-dimensional Brownian dynamics simulations of disks under harmonic potentials. 
	 We plot standard MSDs (circles) according to eq. (\ref{msd}) and relaxed MSDs (squares), eq. (\ref{eq:msd_relax}),  
	 for three different values of the input optical stiffness: $\kappa_o = 0.04$ $\mu$N/m (black), $0.5$ $\mu$N/m (red) and $1.8$ $\mu$N/m (blue).
	 (Inset) Linear behavior of the function $\xi(t)$, eq. (\ref{eq:xi}), for these data allowing to obtain $\lambda$ and the diffusion coefficient $D$.}
	\label{fig:msd-xi}
	\end{centering}
\end{figure}

To have into account this effect, we develop an alternative iterative method to obtain the diffusion coefficient. 
From a step $n$ to a total number of steps $N$, we define a \textit{relaxed} mean-squared displacement, MSD$^*(t)$:
\begin{equation}
  \text{MSD}^*(t) \equiv \langle \Delta x_{n,\lambda}^2 \rangle \equiv \frac{1}{N-n} \sum_i \left(x_{n+i} - x_{i} e^{-\lambda n \tau }\right)^2 \label{eq:msd_relax}
\end{equation}
where this MSD should verify eq. (\ref{alpha}). 
By using here that $\sigma_\kappa^2 \equiv k_B T/\kappa$, we can define
a function $\xi(n\tau)$:
\begin{equation}
  \xi(n\tau) \equiv - \log\left( 1 - \frac{\langle\Delta x_{n,\lambda})^2 \rangle}{\sigma_\kappa^2} \right) = 2\lambda n \tau \label{eq:xi}
\end{equation}
The function $\xi(n\tau)$ grows linearly with time with slope $2 \lambda$, 
something which allows to obtain $\lambda$ from a linear regression. 
By calculating $\kappa$ independently from the distribution $\rho_\kappa (r)$ (\ref{Boltzmann}) (as in Fig. \ref{fig:rho-k}) 
or by the plateau in the MSD, 
and applying an iterative method until obtaining a stable $\lambda$ value.

In Fig. \ref{fig:msd-xi}, we show an example of one-dimensional standard MSD calculation, eq. (\ref{msd}) 
and the relaxed MSD$^*$ defined by eq. (\ref{eq:msd_relax}), 
for several values of the input trap stiffness $\kappa_o$.
Both quantities have a similar behavior, 
displaying a plateau at higher times when reaching the time-scale $\tau_k$. 
However, the MSD$^*$ shows a correction on this plateau, which increases when the restoring force is more intense.
In the Inset of Fig. \ref{fig:msd-xi}, the linear behavior of the defined function $\xi(t)$, eq. (\ref{eq:xi}), 
can be seen, allowing to obtain $\lambda$ and, consequently, $D$.

\begin{table}[htp]
	\begin{centering}	
	\begin{tabular}{cllll}
	\hline
	%Nominal & \multicolumn{3}{|c|}{Experimental} \\
	%\hline
	$\kappa_o$ ($\mu$N/m) & $\kappa$ ($\mu$N/m) & $\lambda$ (s$^{-1}$) & $D^*$ ($\mu\mathrm{m}^2$/s) &$D$ ($\mu\mathrm{m}^2$/s)\\
	\hline
	$0.02$ & $0.020 \pm 0.002$ & $1.2 \pm 0.2$ & $0.24 \pm 0.02$ & $0.247 \pm 0.005$  \\ 
	$0.04$ & $0.052 \pm 0.006$ & $3.1 \pm 0.6$ & $0.24 \pm 0.02$ & $0.242 \pm 0.011$ \\
	$0.06$ & $0.052 \pm 0.003$ & $2.97 \pm 0.03$ & $0.232 \pm 0.012$ & $0.239 \pm 0.004$ \\
	$0.08$ & $0.077 \pm 0.005$ & $4.6 \pm 0.3$ & $0.242 \pm 0.002$ & $0.236 \pm 0.006$ \\
	$0.1$ & $0.099 \pm 0.002$ & $5.9 \pm 0.5$ & $0.243 \pm 0.014$ & $0.239 \pm 0.003$ \\
	$0.2$ & $0.19 \pm 0.02$ & $10.4 \pm 1.7$ & $0.223 \pm 0.010$ & $0.227 \pm 0.005$ \\
	$0.3$ & $0.31 \pm 0.03$ & $19 \pm 3$ & $0.25 \pm 0.02$ & $0.226 \pm 0.003$ \\
	$0.4$ & $0.400 \pm 0.002$ & $23.9 \pm 1.3$ & $0.244 \pm 0.012$ & $0.245 \pm 0.002$ \\
	$0.5$ & $0.4887 \pm 0.0010$ & $28.5 \pm 0.3$ & $0.238 \pm 0.003$ & $0.233 \pm 0.004$ \\
	$0.6$ & $0.613 \pm 0.004$ & $37.7 \pm 1.3$ & $0.251 \pm 0.007$ & $0.238 \pm 0.002$ \\
	$0.8$ & $0.801 \pm 0.014$ & $46 \pm 2$ & $0.233 \pm 0.007$ & $0.235 \pm 0.004$ \\ 
	$1.0$ & $0.988 \pm 0.012$ & $60 \pm 4$ & $0.249 \pm 0.012$ & $0.2403 \pm 0.0012$ \\
	$1.2$ & $1.20 \pm 0.04$ & $68 \pm 4$ & $0.231 \pm 0.007$ & $0.238 \pm 0.002$ \\
	$1.4$ & $1.407 \pm 0.004$ & $83.5 \pm 0.4$ & $0.242 \pm 0.002$ & $0.243 \pm 0.002$ \\
	$1.6$ & $1.59 \pm 0.04$ & $96 \pm 4$ & $0.248 \pm 0.004$ & $0.242 \pm 0.004$ \\ 
	$1.8$ & $1.79 \pm 0.03$ & $107 \pm 3$ & $0.243 \pm 0.011$ & $0.240 \pm 0.006$ \\ 
	$2.0$ & $1.96 \pm 0.09$ & $117 \pm 7$ & $0.243 \pm 0.003$ & $0.241 \pm 0.003$ \\
	\hline
	\end{tabular}
	\caption{Results from experiment-like Brownian dynamics simulations by increasing the input trap stiffness $\kappa_o$. 
	We show the obtained $\kappa$ values, $\lambda$ and, finally, the diffusion coefficients $D^*$ obtained from the iterative method.
	The table summarizes the results for $D$ calculated used the standard MSD and eq. (\ref{Chandrasekhar}).
	The averages provide $\langle D^* \rangle =  0.241 \pm 0.008$ $\mu$m$^2$/s, and $\langle D \rangle =  0.238 \pm 0.007$ $\mu$m$^2$/s,  
	while the theoretical value is $D = 0.239$ $\mu$m$^2$/s.}
	\label{tab:D}
	\end{centering}
\end{table}

In Table \ref{tab:D}, we show a complete set of the values for self-diffusion coefficients 
in the range of the trap stiffnesses used in video-microscopy experiments. 
We summarize the values for the self-diffusion coefficients 
calculated using the standard MSD and eq. (\ref{Chandrasekhar}), $D$,
and through the iterative method, $D^*$.
The data shown is an average of the calculations for coordinates $x$ and $y$, 
which have been analyzed separately after simulations.
We also show the $\kappa$ values obtained from the 
simulated data of the position of the disk to be compared with the input data. 
This comparison between $\kappa_o$ and $\kappa$ allows to
understand the reach of the experimental-like simulations, which are statistically limited, 
as it should be when measuring with a standard video-microscopy set-up. 
The coefficient of diffusion values do not depend on the $\kappa$ value, 
allowing to calculate an average value for $D$ from the data of Table \ref{tab:D}.
The averages provide $\langle D^* \rangle =  0.241 \pm 0.008$ $\mu$m$^2$/s 
and $\langle D \rangle =  0.238 \pm 0.007$ $\mu$m$^2$/s.
Both values nicely agree with the theoretical value, with a relative error lower than 1\%.

\section{Conclusions}

We have developed experiment-like Brownian dynamics simulations of two-dimensional disks under harmonic potentials
to evaluate the reach of experiments of microsphere diffusion with optical tweezers, 
which are modeled as external Hookean forces.
These type of computational studies, based on emulating experimental set-ups, 
are quite useful to design time- and cost-efficient experimental procedures.
After summarize the basic theory of Brownian motion applied to the relevant time-scale, 
we observe that the solution of the Fokker-Planck equation to this stochastic system
allows us to modify the standard definition of the mean-square displacement by including a memory term 
in the initial position of the disk's jumps. 
Based on this relaxed MSD, we propose an alternative method to obtain the self-diffusion coefficient.
The values calculated through that method are compared to the self-diffusion coefficients
obtained using the standard mean-square displacement.
Under the experimental conditions, the averaged values of the diffusion coefficients obtained from both methods return values which differ 
from the theoretical less than 1\%.

\section{Acknowledgments}

We want to thanks F. Ortega for joined investigations with optical tweezers, 
J.A. Torre for his computational support and
J.C. G\'{o}mez-S\'{a}ez for her proofreading of the English texts. 
This research has been supported by MINECO by project FIS2013-47350-C5-5-R.

%\begin{figure}[tb]
	%\begin{centering}
	%\includegraphics[width=0.5\textwidth]{logoEC}
	%\caption{EasyChair logo}
	%\label{fig:easychair-logo}
	%\end{centering}
%\end{figure}

%\begin{table}[htp]
%	\begin{centering}
		%\begin{tabular}{lrrrrrrrr}
		%\hline
		%ATP System            & LTB & Avg  &Prfs & SOTA & \multicolumn{1}{c}{$\mu$} & CYC & MZR & SMO \\
		%                      & /100& time & out & Con. & Eff. & /35 & /40 & /25 \\
		%\hline
		%Vampire-LTB 11.0      &  69 & 24.5 &  69 & 0.37 & 28.1 &  23 &  22 &  24 \\
		%iProver-SInE 0.7      &  67 & 76.5 &   0 & 0.36 &  8.8 &  28 &  14 &  25 \\
		%\hline
		%\end{tabular}
		%\caption{LTB division results}
		%\label{tab:ltbexample}
	%\end{centering}
%\end{table}

%------------------------------------------------------------------------------

\bibliographystyle{plain}
%\bibliography{easychair}
%\bibliography{mpancorbo_iccs17_v2} 

%------------------------------------------------------------------------------
% Index
%\printindex

%------------------------------------------------------------------------------
\end{document}